\documentclass[10pt,journal]{IEEEtran}

\usepackage{amsmath,amssymb,amsfonts}
\usepackage{bm}
\usepackage{graphicx}
\usepackage{cite}
\usepackage{comment}
\usepackage{placeins}
\usepackage{multirow}
\begin{document}

\title{Reduced-Observation Approximation of Near-Field Gaussian Covariance Matrices}

\author{
Marco Moretti, \emph{Member, IEEE}
\thanks{M. Moretti is with the Dipartimento di Ingegneria dell’Informazione, University of Pisa, and also with CNIT (e-mail: marco.moretti@unipi.it). This work has been supported by the Italian NRRP of NextGenerationEU, Program ``RESTART'', Cascade Call SMART, and also  by the Italian Ministry of Education and Research  through the FoReLab Project. The author used OpenAI ChatGPT for language editing. All technical content, analyses, results, and conclusions were developed and verified by the author.}
}

\maketitle

\begin{abstract}
Near-field covariance matrices are central to localization, covariance-aware estimation, and linear MMSE filtering in large-aperture arrays. Under Gaussian range–angle uncertainty, spherical-wave propagation leads to a two-dimensional covariance integral whose numerical evaluation becomes computationally demanding for large numbers of antenna elements. This letter develops a reduced-observation framework for near-field Gaussian covariance approximation. Standard Gram-matrix and truncated-SVD representations are used as computational tools to evaluate the dominant covariance spectrum without explicitly diagonalizing the full covariance matrix. However, these tools do not by themselves determine how fine the quadrature grid must be to approximate the unknown continuous covariance to a prescribed dominant-spectrum accuracy. To address this problem, we introduce a reference-free a posteriori spectral-error estimator based only on dominant-spectrum differences across successive grid resolutions. The estimator exploits local nested-grid convergence information and supports  selection of the smallest tested quadrature resolution whose estimated dominant-spectrum error is below a prescribed tolerance without access to the exact continuous covariance. Numerical results assess the resulting convergence tracking and demonstrate substantial complexity reduction for large near-field arrays.
\end{abstract}


\section{Introduction}

Spatial covariance matrices \cite{dong2022near} are central to
array-processing tasks such as localization~\cite{hua2024near},
channel estimation~\cite{demir2024spatial}, beamforming~\cite{7919262},
and linear MMSE filtering~\cite{long2024mmse}. The combination of
high carrier frequencies and electrically large arrays allows hundreds
or thousands of antenna elements to be deployed within practical
apertures, while extending the radiative near-field region over
relevant communication distances \cite{10716601,XL_MIMO_TUTORIAL}.
In this regime, spherical-wave propagation makes the steering response
depend jointly and nonlinearly on range and angle. Under Gaussian
range--angle uncertainty, the resulting covariance generally requires
a two-dimensional numerical integration, whose computational cost grows
rapidly with the number of array elements.

A practical challenge is to select the quadrature resolution without access to the continuous covariance. While reduced Gram-matrix and truncated-SVD representations can lower the cost of evaluating the covariance spectrum at a given resolution, they do not by themselves indicate how fine the quadrature grid must be to achieve a prescribed approximation accuracy. We address this problem through
reference-free estimation of the dominant-spectrum error.

The main contributions are as follows:
\begin{itemize}
    \item We introduce an a posteriori estimator of the
    dominant-spectrum quadrature error based on a local
    Richardson-type model. The convergence order and error magnitude
    are inferred from spectral differences on three nested grids,
    without an externally calibrated order or a high-resolution
    reference.

    \item We combine coarse- and fine-grid estimates in a hierarchical
    criterion governed by the available refinement budget.
    This criterion supports selection of the smallest tested
    resolution whose estimated dominant-spectrum error is below
    a prescribed tolerance.
\end{itemize}

\section{System Model and Covariance Approximation}

We consider a near-field line-of-sight (LoS) channel generated by a
single-antenna source and received by an array composed of $M$
elements with \emph{generic positions} $\{\mathbf p_m\}_{m=1}^M$. 
Under the considered range--angle uncertainty model, the
source position is parameterized by the range--angle vector
$
    \mathbf q =[        r,\theta]^{T}  \in\mathcal D,
$
where $\mathcal D$ denotes the considered two-dimensional near-field region.
For a source located at $\mathbf q$, the near-field LoS steering vector is
denoted by $\mathbf a(\mathbf q)\in\mathbb C^M$, with
\begin{equation}
[\mathbf a(\mathbf q)]_m
=
\exp\left(
-j\frac{2\pi}{\lambda}
\left[
d_m(\mathbf q)-d_0(\mathbf q)
\right]
\right),
\label{eq:steering_los}
\end{equation}
where $\lambda$ is the wavelength,
$d_m(\mathbf q)=\|\mathbf p(\mathbf q)-\mathbf p_m\|_2$ is the propagation
distance to the $m$-th receiving element, and $d_0(\mathbf q)$ is a
reference distance, e.g., the distance to the array center. Unlike the
far-field plane-wave response, these element-dependent distances introduce
a joint nonlinear dependence on range and angle.
Let $f(\mathbf q)$ denote the probability density of the source position
over $\mathcal D$, normalized such that
$\int_{\mathcal D}f(\mathbf q)\,d\mathbf q=1$. The corresponding near-field
spatial covariance matrix\footnote{A phase-only response is used for notational simplicity. The
formulation also applies to near-field steering vectors including
element-dependent amplitudes.} is
\begin{equation}
\mathbf R
=
\int_{\mathcal D}
\mathbf a(\mathbf q)
\mathbf a^H(\mathbf q)
f(\mathbf q)\,
d\mathbf q .
\label{eq:R_continuous}
\end{equation}

We focus on Gaussian range--angle uncertainty. Specifically,
$f(\mathbf q)$ is centered at the nominal source position
$\boldsymbol{\mu}_q=[r_0,\theta_0]^T$ and is restricted to $\mathcal D$:
\begin{equation}
f(\mathbf q)
=
\frac{1}{\eta}
\exp\left(
-\frac{1}{2}
(\mathbf q-\boldsymbol{\mu}_q)^T
\boldsymbol{\Sigma}_q^{-1}
(\mathbf q-\boldsymbol{\mu}_q)
\right),
\qquad
\mathbf q\in\mathcal D ,
\label{eq:gaussian_source_density}
\end{equation}
where $\eta$ normalizes the density over $\mathcal D$. We assume
independent radial and angular uncertainties, i.e., 
$
\boldsymbol{\Sigma}_q
=
\operatorname{diag}(\sigma_r^2,\sigma_\theta^2).
$
Because the near-field response depends nonlinearly on both coordinates,
\eqref{eq:R_continuous} generally requires numerical integration.
We consider two-dimensional tensor-product quadrature grids with $N$
points per dimension, corresponding to $Q=N^2$ quadrature nodes
$\{\mathbf q_i\}_{i=1}^{Q}$ and weights $\{w_i\}_{i=1}^{Q}$, including
the quadrature measure and $f(\mathbf q_i)$. The covariance approximation
at resolution $N$ is then
\begin{equation}
\mathbf R_N
=
\sum_{i=1}^{Q}
w_i
\mathbf a(\mathbf q_i)
\mathbf a^H(\mathbf q_i),
\qquad Q=N^2 .
\label{eq:R_quadrature}
\end{equation}

\section{Reduced Observation Representation}

Let
\(
\mathbf a_i\triangleq\mathbf a(\mathbf q_i)
\)
denote the near-field steering vector evaluated at the $i$-th quadrature
node, and define
\[
\mathbf H_{N}
=
\left[
\sqrt{w_1}\mathbf a_1,\,
\sqrt{w_2}\mathbf a_2,\,
\ldots,\,
\sqrt{w_{N^{2}}}\mathbf a_{N^{2}}
\right]
\in\mathbb C^{M\times Q}.
\]
Then, the quadrature covariance in \eqref{eq:R_quadrature} can be written as
\begin{equation}
    \mathbf R_N=\mathbf H_{N}\mathbf H_{N}^H .
    \label{eq:R_HH}
\end{equation}
When $Q<M$, and especially for $Q\ll M$, spectral computation can be moved
from the $M\times M$ covariance matrix to the reduced Gram matrix
\begin{equation}
    \mathbf G_{N}
    =
    \mathbf H_{N}^H\mathbf H_{N}
    \in\mathbb C^{Q\times Q}.
    \label{eq:G}
\end{equation}

Since $\mathbf H_{N}\mathbf H_{N}^H$ and $\mathbf H_{N}^H\mathbf H_{N}$ have the same
nonzero eigenvalues,
\begin{equation}
    \lambda_k^{(N)}
    =
    \lambda_k(\mathbf G_{N}),
    \qquad
    k=1,\ldots,\operatorname{rank}(\mathbf H_{N}).
    \label{eq:gram_eig_relation}
\end{equation}
Thus, the dominant covariance spectrum required by the spectral-error
estimator can be obtained without diagonalizing $\mathbf R_N$ in the full
array domain.
Further computational savings are possible when the covariance spectrum is
compressible and only the first $K$ dominant modes are required. In this
case, a rank-$K$ truncated SVD of $\mathbf H_{N}$,
\begin{equation}
    \mathbf H_{N}
    \approx
    \mathbf U_{N,K}
    \boldsymbol{\Sigma}_{N,K}
    \mathbf V_{N,K}^H,
    \label{eq:H_truncated_svd}
\end{equation}
directly provides both the dominant covariance eigenvalues and the
corresponding covariance-domain modes. 
Specifically,
\begin{equation}
    \mathbf R_N
    =
    \mathbf H_N\mathbf H_N^H
    \approx
    \mathbf U_{N,K}
    \boldsymbol{\Sigma}_{N,K}^2
    \mathbf U_{N,K}^H
    \triangleq
    \mathbf R_{N,K},
    \label{eq:R_truncated}
\end{equation}
where the dominant covariance eigenvalues satisfy
\[
    \lambda_k^{(N)}
    =
    \lambda_k(\mathbf G_N)
    =
    \sigma_k^2(\mathbf H_N),
    \qquad k=1,\ldots,K.
\]

%**********************
\subsection{Computational Complexity}
\label{subsec:complexity}
A direct
covariance-domain implementation requires forming
\(\mathbf R_N\) as in \eqref{eq:R_HH}, with cost
scaling as \(O(M^2Q)\), followed by the eigendecomposition of an
\(M\times M\) matrix, with cost \(O(M^3)\). Its overall complexity is
therefore
\begin{equation}
    C_{\rm full}(Q) = O(M^2Q+M^3).
    \label{eq:complexity_full}
\end{equation}

The reduced Gram formulation avoids the full covariance-domain
eigendecomposition. It forms
\(\mathbf G_{N}\in\mathbb C^{Q\times Q}\), with cost
\(O(MQ^2)\), and diagonalizes the resulting \(Q\times Q\) matrix, with
cost \(O(Q^3)\). Hence,
\begin{equation}
    C_{\rm Gram}(Q) = O(MQ^2+Q^3).
    \label{eq:complexity_gram}
\end{equation}
This approach is advantageous when \(Q<M\), since the eigendecomposition
is moved from the array domain to the quadrature domain.

When only the first \(K\) dominant covariance modes are needed, a
rank-\(K\) truncated SVD of \(\mathbf H_{N}\) directly provides both the
dominant covariance eigenvalues and the corresponding covariance-domain
modes. Following standard large-scale SVD implementations, the approximated 
cost scales as
\begin{equation}
    C_{\rm tSVD}(Q,K)=O(MQK),
    \label{eq:complexity_tsvd}
\end{equation}
up to lower-order orthogonalization and projection terms
\cite{halko2011finding}. This regime is particularly relevant for the
near-field LoS covariance matrices considered here, whose spectrum is
typically concentrated in a limited number of dominant modes under
localized range--angle uncertainty \cite{besson2000decoupled,cui2022channel}.
Materializing this factorization as a dense \(M\times M\) matrix would add
\(O(M^2K)\) operations, whereas these costs are
avoided when the low-rank factors are used directly.

%************
\section{Dominant-Spectrum Error Metric}

A natural way to measure the spectral approximation error is to
normalize the eigenvalue perturbation by the total covariance power.
Accordingly, we define
\begin{equation}
e(N)
=
\frac{
\left\|
\boldsymbol{\lambda}_{1:M}^{(N)}
-
\boldsymbol{\lambda}_{1:M}
\right\|_2
}{
\sum_{k=1}^{M}\lambda_k
}.
\label{eq:full_normalized_error}
\end{equation}

In near-field LoS scenarios under localized range--angle uncertainty,
the covariance spectrum is typically compressible
\cite{zeng2021degrees}, with most of the covariance energy concentrated
in $K<M$ dominant modes. Accordingly, for a prescribed number $K$ of
dominant modes, let
$\boldsymbol{\lambda}_{1:K}
=[\lambda_1,\ldots,\lambda_K]^T$
and
$\boldsymbol{\lambda}_{1:K}^{(N)}
=[\lambda_1^{(N)},\ldots,\lambda_K^{(N)}]^T$.
We then define the corresponding trace-normalized dominant-spectrum
error as
\begin{equation}
\widetilde e_K(N)
=
\frac{
\left\|
\boldsymbol{\lambda}_{1:K}^{(N)}
-
\boldsymbol{\lambda}_{1:K}
\right\|_2
}{
\operatorname{tr}(\mathbf R)
}.
\label{eq:trace_normalized_error}
\end{equation}

For the phase-only steering model in \eqref{eq:steering_los},
each steering vector satisfies
$\|\mathbf a(\mathbf q)\|_2^2=M$. Since $f(\mathbf q)$ is normalized
over $\mathcal D$, linearity of the trace gives
\begin{equation}
\operatorname{tr}(\mathbf R)
=
\int_{\mathcal D}
\|\mathbf a(\mathbf q)\|_2^2
f(\mathbf q)\,d\mathbf q
=
M .
\label{eq:trace_R}
\end{equation}
Hence, the normalization factor in
\eqref{eq:trace_normalized_error} is known a priori for the considered
model and does not require the exact covariance eigenvalues.
However, evaluating $\widetilde e_K(N)$ still requires the exact
dominant covariance spectrum
$\boldsymbol{\lambda}_{1:K}$, which is unavailable in practice.
This motivates the reference-free a posteriori estimator developed
next.
 
%*******************

\section{Spectral-Error Estimator}
\label{sec:spectral_estimator}

We define the normalized grid spacing as
$\rho(N)=1/(N-1)$.
For two resolutions $N_a<N_b$, we define the observable
trace-normalized spectral difference
\begin{equation}
    \Delta_K(N_a,N_b)
    =
    \frac{
    \left\|
    \boldsymbol{\lambda}_{1:K}^{(N_b)}
    -
    \boldsymbol{\lambda}_{1:K}^{(N_a)}
    \right\|_2
    }{M}.
    \label{eq:pairwise_spectral_difference}
\end{equation}
Unlike $\widetilde e_K$, this quantity does not require knowledge of the
exact covariance spectrum.
Following the leading-order error model underlying Richardson
extrapolation~\cite{zlatev2017richardson}, we assume locally that
\begin{equation}
    \boldsymbol{\lambda}_{1:K}^{(N)}
    -
    \boldsymbol{\lambda}_{1:K}
    \simeq
    \mathbf c\,\rho(N)^p ,
    \label{eq:local_power_model}
\end{equation}
where \(p>0\) is the local convergence order. Consequently,
\begin{equation}
    \widetilde e_K(N)
    \simeq
    C\rho(N)^p,
    \qquad
    C=\frac{\|\mathbf c\|_2}{M}.
    \label{eq:scalar_power_model}
\end{equation}
Both the scale factor \(C\) and the convergence order \(p\) are unknown,
so that at least two independent spectral differences are needed to
estimate them.

Consider three \emph{nested} grids $N_{c},N_{f},N_{ff}$ with
    $N_f=2N_c-1$  and 
    $N_{ff}=2N_f-1$,
so that the number of intervals is doubled at each refinement. The corresponding grid spacing is $\rho_{c} =\rho(N_c)$, $\rho_{f}=\rho(N_f)=2^{-1}\rho_{c}$, and $\rho_{ff}=\rho(N_{ff})=2^{-1}\rho_f=2^{-2}\rho_c$.
Defining
$    \Delta_{cf}=\Delta_K(N_c,N_f)$ and $\Delta_{ff}=\Delta_K(N_f,N_{ff})$,
the model in \eqref{eq:scalar_power_model}  gives 
\begin{equation}
\frac{\Delta_{cf}}{\Delta_{ff}}
\simeq
\frac{\rho_c^p-\rho_f^p}
     {\rho_f^p-\rho_{ff}^p}
=2^p .
\end{equation}
and the local convergence order is estimated as
\begin{equation}
    \widehat p
    =
    \log_{2}(\Delta_{cf}/\Delta_{ff}).
    \label{eq:local_order_estimate}
\end{equation}
The scale factor is  estimated from either spectral difference as
\begin{equation}
    \widehat C
    =
    \frac{\Delta_{cf}}
    {\rho_c^{\widehat p}-\rho_f^{\widehat p}}
    =
    \frac{\Delta_{ff}}
    {\rho_f^{\widehat p}-\rho_{ff}^{\widehat p}}.
    \label{eq:local_scale_estimate}
\end{equation}

Substituting \eqref{eq:local_scale_estimate} and
\eqref{eq:local_order_estimate} into \eqref{eq:scalar_power_model}
yields the local estimates at $N_c$ and $N_f$. For fixed
$K$, we omit the subscript $K$ and denote them by
$\widehat e_{\rm c}$, and $\widehat e_{\rm f}$
\begin{equation}
    \widehat e_{\rm c} = \frac{\Delta_{cf}^{2}}{\Delta_{cf}-\Delta_{ff}}, 
    \widehat e_{\rm f} = \frac{\Delta_{cf}\Delta_{ff}}{\Delta_{cf}-\Delta_{ff}}.
    \label{eq:local_error_estimator}
\end{equation}
Because adjacent nested triplets may overlap, the same grid resolution
can generally be associated with more than one a posteriori error
estimate. For example, $N=3$ is the fine grid of the triplet $(2,3,5)$
and the coarse grid of the subsequent triplet $(3,5,9)$, thus providing
both $\widehat e_{\rm f}(3)$ and $\widehat e_{\rm c}(3)$.
To motivate an operational selection rule, we examine the local
sensitivity of these estimates by introducing the observable local
contraction factor
\begin{equation}
    r_\Delta
    \triangleq
    \frac{\Delta_{ff}}{\Delta_{cf}}
    = 2^{-\widehat p},
    \qquad 0<r_\Delta<1.
    \label{eq:contraction_factor}
\end{equation}
In terms of $r_\Delta$, the coarse- and fine-grid estimates become
\begin{equation}
    \widehat e_{\rm c}
    =\frac{\Delta_{cf}}{1-r_\Delta},
    \qquad
    \widehat e_{\rm f}
    =\frac{\Delta_{ff}}{1-r_\Delta}.
    \label{eq:contraction_error_estimates}
\end{equation}
Triplets not satisfying $0<r_\Delta<1$ are discarded.
For fixed $\Delta_{cf}$, define the
relative sensitivity to $r_\Delta$ as
\begin{equation}
    S_x
    =
    \left|
    \frac{\partial \log \widehat e_x}
         {\partial \log r_\Delta}
    \right|.
\end{equation}
From \eqref{eq:contraction_error_estimates},
\begin{equation}
    S_{\rm c}
    =\frac{r_\Delta}{1-r_\Delta},
    \qquad
    S_{\rm f}
    =\frac{1}{1-r_\Delta},
    \qquad
    S_{\rm c}=r_\Delta S_{\rm f}<S_{\rm f}.
    \label{eq:sensitivity}
\end{equation}
For a fixed triplet and fixed $\Delta_{cf}$, the coarse-grid estimate
has lower relative sensitivity to $r_\Delta$. This comparison does
not establish a general sensitivity ranking across different triplets.
For $r_\Delta\ll1$,
\[
    \widehat e_{\rm f}=\frac{r_\Delta\Delta_{cf}}{1-r_\Delta}
    =\frac{\Delta_{ff}}{1-r_\Delta}
    \simeq \Delta_{ff}.
\]
For two overlapping triplets $(N_0,N_1,N_2)$ and $(N_1,N_2,N_3)$,
let $\Delta_i=\Delta_K(N_i,N_{i+1})$, $i=0,1,2$, and
$r_0=\Delta_1/\Delta_0$, $r_1=\Delta_2/\Delta_1$.
Applying \eqref{eq:contraction_error_estimates} to the common
resolution $N_1$ gives
\[
    \frac{\widehat e_{\rm f}(N_1)}
         {\widehat e_{\rm c}(N_1)}
    =
    \frac{1-r_1}{1-r_0}.
\]
Hence, if $r_1\simeq r_0$ the two estimates nearly coincide, while for
$r_0,r_1\ll1$ both approach $\Delta_1$.
%
%
%For two overlapping triplets $(N_0,N_1,N_2)$ and $(N_1,N_2,N_3)$,
%define
%$
%\Delta_i=\Delta_K(N_i,N_{i+1})$, ($i=0,1,2$).
%The fine-grid estimate from the first triplet and the coarse-grid
%estimate from the second, both associated with the  resolution
%$N_1$, are
%\[
%    \widehat e_{\rm f}(N_1)
%    =\frac{\Delta_1}{1-r_0},
%    \qquad
%    \widehat e_{\rm c}(N_1)
%    =\frac{\Delta_1}{1-r_1},
%\]
%with
%$r_0=\Delta_1/\Delta_0$ and $r_1=\Delta_2/\Delta_1$.
%Hence, if $r_1\simeq r_0$ they nearly coincide, and for
%$r_0,r_1\ll1$,
%\[
%    \widehat e_{\rm f}^{(0)}(N_1)
%    \simeq
%    \widehat e_{\rm c}^{(1)}(N_1)
%    \simeq \Delta_1 .
%\]

Let $N_{\max}$ denote the largest grid size allowed by the prescribed
quadrature budget. A coarse-grid estimate at $N_c$ requires the two
subsequent nested refinements up to
$N_{ff}=4N_c-3$, whereas a fine-grid estimate at $N_f$ only requires
$N_{ff}=2N_f-1$. Thus,
\begin{equation}
    N_c\leq\frac{N_{\max}+3}{4},
    \qquad
    N_f\leq\frac{N_{\max}+1}{2}.
    \label{eq:budget_ranges}
\end{equation}
Motivated by this local sensitivity analysis, we adopt an operational
hierarchy that prioritizes valid coarse-grid estimates within the
prescribed budget and uses valid fine-grid estimates when
coarse-grid estimates are unavailable.

Let $\mathcal N_c$ and $\mathcal N_f$ denote the sets of
grid resolutions for which a valid coarse- and fine-grid estimate,
respectively, are available within the prescribed budget.
We define the
hierarchical estimator
\begin{equation}
\widehat e_{\mathrm{H},K}(N)
=
\begin{cases}
\widehat e_{\rm c}(N), & N\in\mathcal N_c,\\[1mm]
\widehat e_{\rm f}(N), & N\in\mathcal N_f\setminus\mathcal N_c.
\end{cases}
\label{eq:hierarchical_estimator}
\end{equation}

The resulting estimator is entirely reference-free and requires neither
an externally calibrated convergence order nor a high-resolution
reference. It is intended as an operational a posteriori estimate of the
dominant-spectrum quadrature error rather than as a universal analytical
upper bound.
For a target tolerance $\epsilon$, $\widehat e_{\mathrm{H},K}(N)$ provides
an a posteriori criterion for selecting the smallest tested resolution whose estimated error is below the target tolerance.
%*******************

\section{Numerical Results}

We consider a $128\times16$ UPA with half-wavelength spacing at
$f_c=28$ GHz, yielding $M=2048$ antennas. 
The source is located at a nominal distance $r_0=10$ m and angle
$\theta_0=5^\circ$, with a fixed vertical offset of $1.5$ m.
The angular spread is $\sigma_\theta=5^\circ$, while
$\sigma_r=r_0\tan(\sigma_\theta)$. The integration domain is truncated to
$\pm4$ standard deviations in both dimensions. We retain the
$K=50$ dominant covariance modes. 
A dense grid with $N_{\rm ref}=257$ is used only to obtain the
high-resolution reference covariance $\mathbf R_{\rm ref}$ and the
corresponding rank-$K$ quantities
$\mathbf R_{\rm ref,K}$ and
$\boldsymbol\lambda_{1:K}^{(\rm ref)}$ for validation.

\begin{figure}[t]
    \centering
    \includegraphics[width=0.75\columnwidth]{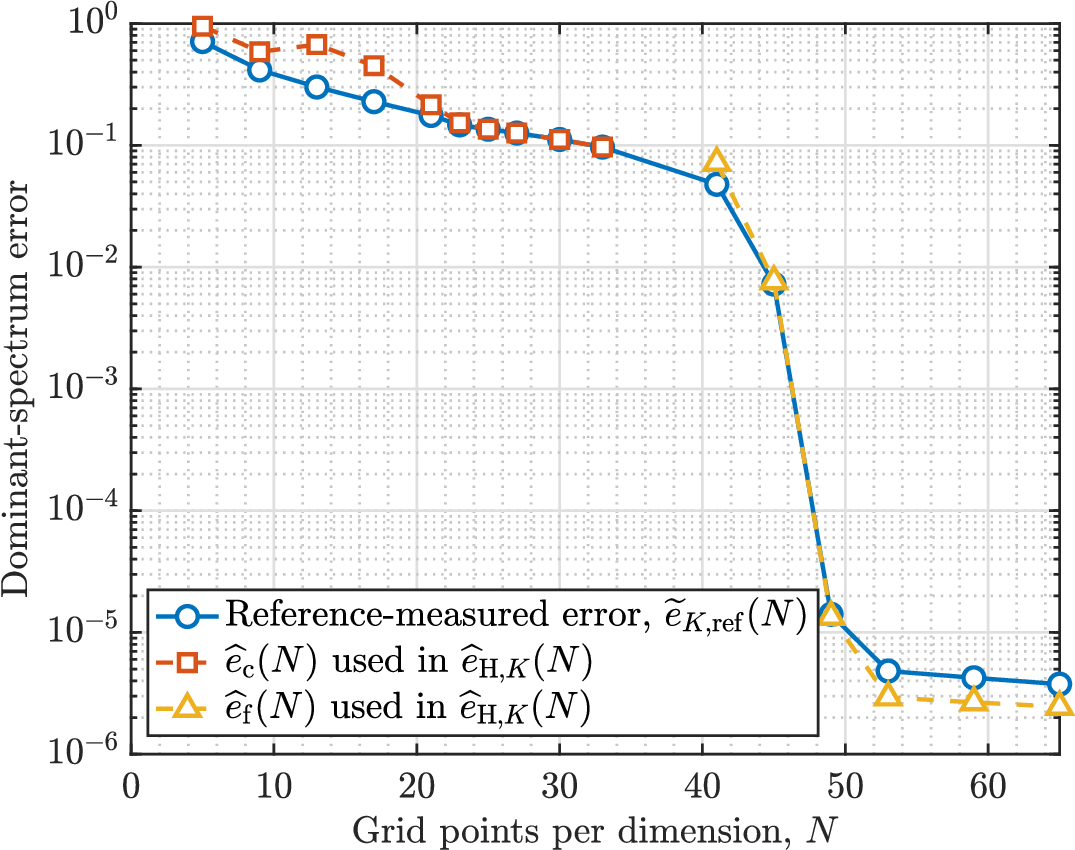}
    \caption{Reference-measured and estimated dominant-spectrum error
versus the quadrature-grid resolution $N$.}
    \label{fig:predictor_validation}
\end{figure}

Fig.~\ref{fig:predictor_validation} compares the reference-measured
dominant-spectrum error $\widetilde e_{K,\mathrm{ref}}(N)$ with the
hierarchical estimate $\widehat e_{\mathrm{H},K}(N)$. The estimator
tracks the error scale over the whole considered range and,  while at coarse resolutions the estimate is
conservative, it becomes
increasingly accurate as the quadrature enters the contracting regime.
 The corresponding decrease of $r_\Delta$ in the fine-grid region is
reported in Table~\ref{tab:lmmse}. 
The proposed estimator is neither an upper nor a lower bound and may therefore lie
on either side of the reference error; nevertheless, results show that it closely tracks
its magnitude and provides an accurate indication of the attained error
level.

\begin{figure}[h]
    \centering
    \includegraphics[width=0.75\columnwidth]{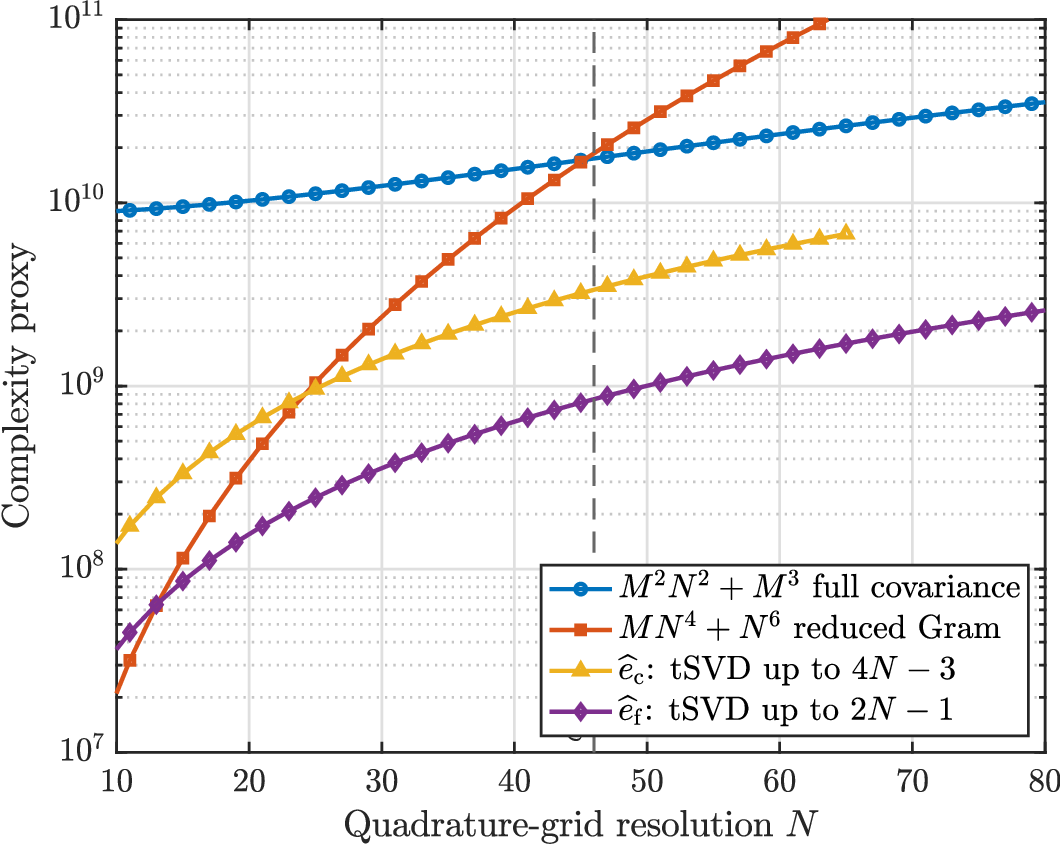}
\caption{Computational complexity versus the quadrature-grid
resolution $N$.}
   \label{fig:compl}
\end{figure}

Fig.~\ref{fig:compl} compares the complexity scalings derived in
Section~\ref{subsec:complexity}. 
The Gram formulation is advantageous for $Q<M$, whereas
truncated SVD reduces the leading computational cost for
$K\ll\min\{M,Q\}$, including grids with $Q\ge M$. The figure also reports the maximum
truncated-SVD complexity required by the coarse- and fine-grid error
estimates. For an error estimate at resolution $N$, these require
look-ahead up to $4N-3$ and $2N-1$, respectively, explaining the
larger admissible resolution range of the fine-grid estimate under a
fixed quadrature budget.

\begin{figure}[b]
    \centering
    \includegraphics[width=0.75\columnwidth]{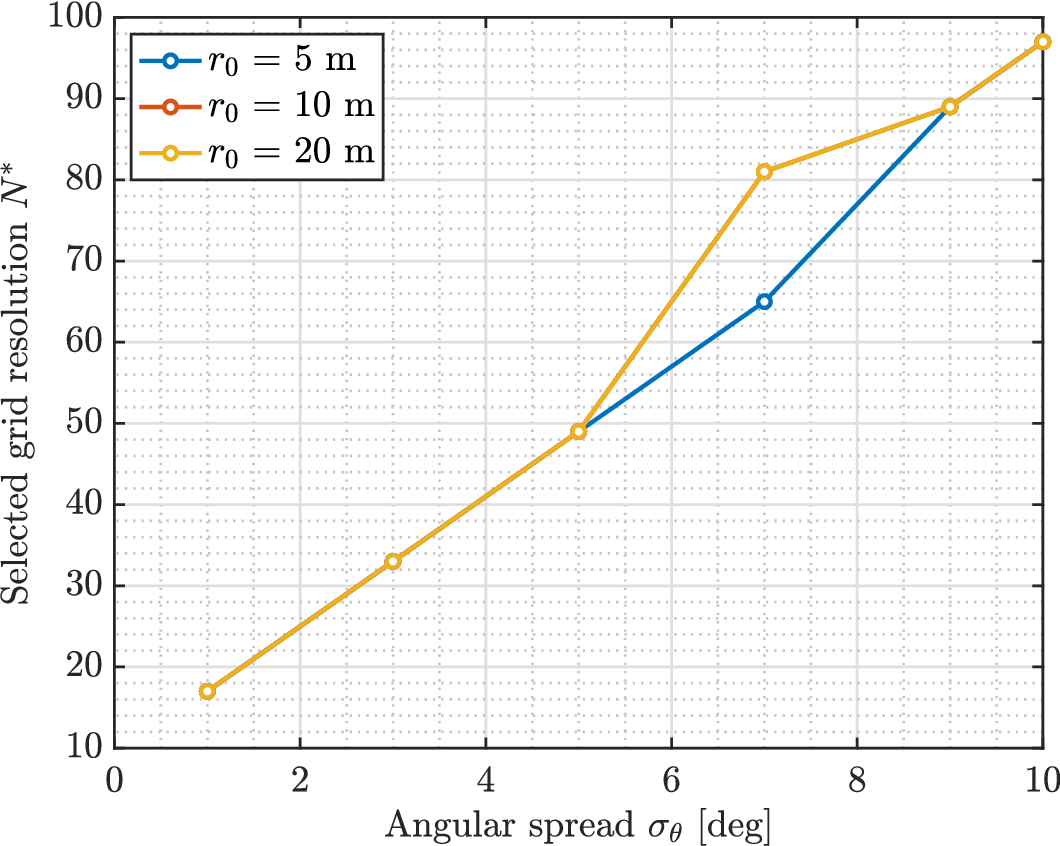}
    \caption{Quadrature-grid resolution $N^\star$ selected by the
hierarchical dominant-spectrum error criterion versus angular spread
$\sigma_\theta$, for different nominal source distances. The target
estimated error is $\epsilon=5\times10^{-5}$ and
$K =100$.}
\label{fig:gridNum}
\end{figure}

Fig.~\ref{fig:gridNum} shows the quadrature-grid resolution selected
by the proposed a posteriori criterion for a target estimated error
$\epsilon=5\times10^{-5}$. We set $K=100$ to account for the increasing
spread of covariance energy across modes at larger angular spreads and
to limit the effect of spectral truncation on grid selection.  The selected resolution increases
systematically with the angular spread, from $N^\star=17$ at
$\sigma_\theta=1^\circ$ to $N^\star=97$ at
$\sigma_\theta=10^\circ$. This behavior reflects the progressively
finer sampling required to resolve the stronger angular variation of
the near-field steering response. The dependence on the nominal
distance is comparatively weak.

To assess the practical impact of the covariance approximation, we
consider
$\mathbf h\sim\mathcal{CN}(\mathbf 0,\mathbf R_{\mathrm{ref},K})$ and
$\mathbf y=\sqrt{\gamma}\mathbf h+\mathbf n$, with
$\mathbf n\sim\mathcal{CN}(\mathbf 0,\mathbf I)$. The average receive SNR,
$\gamma\operatorname{tr}(\mathbf R_{\mathrm{ref},K})/M$, is set to
$10$ dB. The receiver uses the rank-$K$ LMMSE filter
\[
\mathbf W_N
=
\mathbf U_{N,K}
\operatorname{diag}\!\left(
\frac{\sqrt{\gamma}\lambda_k^{(N)}}
     {1+\gamma\lambda_k^{(N)}}
\right)_{k=1}^{K}
\mathbf U_{N,K}^H ,
\]
and we report the mismatch loss
\[
L_N=10\log_{10}\frac{\mathrm{MSE}_N}{\mathrm{MSE}_{\mathrm{ref}}},
\]
where $\mathrm{MSE}_N$ and $\mathrm{MSE}_{\mathrm{ref}}$ denote the
mismatched and matched LMMSE errors, respectively. As shown in
Table~\ref{tab:lmmse}, improved covariance accuracy rapidly reduces the
estimation penalty.

\begin{table}[t]
\centering
\caption{Impact of the covariance approximation on LMMSE channel estimation at $10$ dB SNR.}
\label{tab:lmmse}
\scriptsize
\setlength{\tabcolsep}{4pt}
\begin{tabular}{c|cccc}
\hline
$N$ &
$\widetilde e_{K,\mathrm{ref}}(N)$ &
$\widehat e_{\mathrm{H},K}(N)$ & $r_{\Delta}$ &
$L_{N}$ [dB] \\
\hline
41 & $4.8\times10^{-2}$ & $7.2\times10^{-2}$ & $3.34\times10^{-1}$ & $15.6$ \\
45 & $7.3\times10^{-3}$ & $7.6\times10^{-3}$ & $4.98\times10^{-2}$ & $4.7$ \\
49 & $1.3\times10^{-5}$ & $1.3\times10^{-5}$ & $9.93\times10^{-5}$ & $2.0\times10^{-3}$ \\
\hline
\end{tabular}
\end{table}


\section{Conclusion}
This letter proposed a reduced-observation framework for near-field
Gaussian covariance approximation. The formulation extracts the dominant
covariance spectrum from \(\mathbf R_N=\mathbf H_{N}\mathbf H_{N}^H\) without full
covariance-domain eigendecomposition, and introduces a reference-free a posteriori spectral-error estimator based on grid-to-grid spectral
differences. 
Numerical results confirmed accurate error-scale tracking and substantial
complexity savings for large-aperture near-field arrays, while the LMMSE
example illustrated how improved covariance accuracy translates into a
rapid reduction of estimation loss.
\bibliographystyle{IEEEtran}
\bibliography{references}

\end{document}